\documentclass[a4paper,UKenglish]{lipics-v2021}

\title{Decidability of Elementary Modal Logics is Undecidable}

\author{Jakub Michaliszyn}{University of Wroc\l{}aw}{jmi@cs.uni.wroc.pl}{0000-0002-5053-0347}{}

\authorrunning{Jakub Michaliszyn} %TODO mandatory. First: Use abbreviated first/middle names. Second (only in severe cases): Use first author plus 'et al.'

\Copyright{Jakub Michaliszyn} %TODO mandatory, please use full first names. LIPIcs license is "CC-BY";  http://creativecommons.org/licenses/by/3.0/


\category{}


\relatedversion{} %optional, e.g. full version hosted on arXiv, HAL, or other respository/website
\EventEditors{}
\EventNoEds{2}
\EventLongTitle{}
\EventShortTitle{}
\EventAcronym{}
\EventYear{}
\EventDate{}
\EventLocation{}
\EventLogo{}
\SeriesVolume{42}
\ArticleNo{40}
\usepackage{tcolorbox}
\usepackage{tikz}
\usetikzlibrary{patterns,decorations.pathreplacing,shapes,arrows,calc}
\usetikzlibrary{arrows.meta}

\usepackage[T1]{fontenc}

\newcommand{\str}[1]{{\mathfrak{#1}}}

\newcommand{\N}{{\mathbb N}}
\newcommand{\naw}[1]{{\langle #1\rangle}}

\newcommand{\SAT}{\operatorname{SAT}}
\newcommand{\Own}{\operatorname{Own}}
\newcommand{\tp}{\operatorname{tp}}
\newcommand{\One}{\operatorname{One}}
\newcommand{\ST}{\operatorname{ST}}
\newcommand{\Dom}{\mathcal D}
\newcommand{\Tiles}{\mathcal T}

\newcommand{\klasa}{{\cal K}}
\newcommand{\kp}{\klasa_\Phi}

\newcommand{\sat}[1]{$#1$-SAT${}$}

\newtcolorbox{mybox}[1]{colback=yellow!5!white,colframe=gray!60!black,fonttitle=\small\bfseries,left=2pt,right=2pt,title=#1}

\newcommand{\problem}[3]{
\begin{mybox}{#1}
	\textbf{Input}:  #2    ~\\
	\textbf{Question}: #3 
\end{mybox}
}
\nolinenumbers
\hideLIPIcs

\begin{document}

\maketitle

\begin{abstract} 
We study the decidability of modal satisfiability over classes of Kripke frames defined by universal first-order sentences. Although some such classes yield undecidable satisfiability problems, decidability is guaranteed for wide families of frame conditions, including universal Horn sentences and universal sentences implying transitivity. 
A more than decade-old open problem \cite{MO12,MOK2015} asks a natural question: can decidable and undecidable cases be distinguished algorithmically? 
We prove that this is impossible, already for equality-free universal frame conditions over a single binary relation. More precisely, there is no algorithm that, given such a condition $\Psi$, decides whether its associated local modal satisfiability problem is decidable. We first establish the result for conditions with equality and then eliminate equality by passing to quotients of worlds with identical incoming and outgoing edges.
\end{abstract}

\newpage 

\section{Introduction}

Modal logic is a formalism that allows for describing how truth depends on a state, a
possible situation, or the information available to an agent.
In computer science, its applications
range from reasoning about hardware and software to the analysis of security
protocols and the representation of knowledge. 

Modal logic extends propositional logic  with operators  $\Diamond$ and $\Box$. 
A \emph{frame} consists of a set of worlds and an accessibility relation, and a model
adds a valuation assigning a set of propositional variables to each world.
At a world $w$, the formula $\Diamond\varphi$ holds when some world
accessible from $w$ satisfies $\varphi$. Its dual $\Box\varphi$ holds when
every such world satisfies $\varphi$. 

Many frame classes of interest can be
specified in first-order logic. Given a sentence $\Phi$ over the
signature containing a binary relation symbol $R$, write $\mathcal{K}_\Phi$
for the class of frames satisfying $\Phi$. For instance,
$
  \forall x\,y\,z\,
  \bigl(R(x,y)\wedge R(y,z)\rightarrow R(x,z)\bigr)
$
defines the transitive frames. Such first-order conditions define
elementary classes of frames, and the modal logics determined by these
classes are called elementary modal logics.

This approach generalises many standard modal systems and suggests a
general decision problem: which restrictions on a first-order frame
condition guarantee decidability of the associated modal logic? 
It is convenient to formulate decision problems in terms of satisfiability.
Local satisfiability asks for a
model and a world at which the input formula holds; global satisfiability
requires a model in which it holds at every world. 

Unrestricted first-order frame conditions can lead to undecidability, even for universal formulas.
Hemaspaandra~\cite{H96} showed a universal condition
with equality for which global satisfiability was undecidable.
Subsequently undecidability
of local satisfiability for an equality-free universal condition was obtained~\cite{HS11}. Their
construction used nine variables.

These results raised the question of how many variables were needed in the
frame conditions. Reflexivity, symmetry, transitivity, and the Euclidean
property can all be expressed with at most three variables, and the
standard modal systems associated with these conditions have decidable
satisfiability problems~\cite{ladner}. However, the analysis presented in
\cite{MOK2015} showed that the three-variable setting already admitted
undecidability. The authors exhibited equality-free universal sentences
with three variables yielding undecidable global and local satisfiability
problems, respectively. The constructions also established undecidability
of the corresponding finite-model problems. In the local case, a single
universal clause sufficed.

The variable bound was optimal. The standard translation of modal logic
uses only two first-order variables, reused under quantification.
Consequently, adding a two-variable frame sentence still permits a
reduction to the decidable two-variable fragment~\cite{Mortimer}. This
translation also offers a route to further positive results whenever the
translated modal formula and the frame constraints together belong to a
decidable first-order fragment, for example involving counting~\cite{Har05} or guarded-negation
formalisms~\cite{BtCS}.

A different restriction is suggested by the shape of the standard frame
conditions~\cite{HS08}: they can be written as universal Horn sentences (conjunction of clauses, each with at most one positive
literal). Decidability for all universal Horn frame conditions was 
shown in~\cite{MOK2015} together with a precise complexity characterisation. Apart from inconsistent
conditions, local
satisfiability was classified as either $\mathsf{NP}$-complete or
$\mathsf{PSPACE}$-complete. For global satisfiability, the possible
complexities were $\mathsf{NP}$-complete, $\mathsf{PSPACE}$-complete,
and $\mathsf{EXPTIME}$-complete. 

Another broad decidable family arises by requiring transitivity ~\cite{MO13}. It was shown that modal satisfiability
is decidable over every class of transitive frames defined
by a universal first-order sentence, without imposing
the Horn restriction. For each fixed frame condition,
both local satisfiability and finite local satisfiability
belong to $\mathsf{NEXPTIME}$, while global satisfiability
and finite global satisfiability belong to $\mathsf{NP}$.

Finite satisfiability requires separate treatment, since universal
Horn frame conditions need not have the finite model property.
Nevertheless, both local and global finite satisfiability belong
to $\mathsf{NEXPTIME}$ for every fixed universal Horn
sentence~\cite{MOK2015}.
Beyond the Horn setting, Michaliszyn, Otop, and Witkowski~\cite{MOW2018}
gave a universal frame condition for which global satisfiability is
decidable but finite global satisfiability is undecidable, as well as
conditions with the reverse behaviour.

Another direction is to allow several accessibility relations. In a
multimodal language, each pair $\Diamond_i,\Box_i$ is interpreted using
its own relation $R_i$. Some decision procedures continue to apply in
this setting through the standard translation. Examples arise from
decidable two-variable formalisms with suitable restrictions on
distinguished relations, including variants with two equivalence
relations~\cite{KO05,KMI12}, and from appropriate guarded fragments.
Such results depend on the interaction permitted between the relations,
not merely on the availability of several modal operators.

Michaliszyn, Otop, and Kiero\'nski~\cite{MOK2015} proved that the
unimodal universal Horn decidability theorem did not extend to
arbitrary bimodal frame conditions. They exhibited universal Horn
sentences coupling two accessibility relations for which local or
global satisfiability was undecidable, including the corresponding
finite satisfiability problems. They also conjectured that decidability
could be retained if every clause mentioned only one accessibility
relation. Both the negative result and this conjecture belong to that
earlier work; the restriction in the conjecture concerned interaction
between relations.

In a subsequent preprint, Michaliszyn~\cite{Michaliszyn2018}
showed that universal Horn conditions coupling two modalities could
yield undecidable satisfiability even when one or both accessibility
relations were required to be transitive. Thus transitivity of the
individual relations did not restore the positive unimodal results
in the presence of cross-relation constraints.

More recently, positive results have also been obtained beyond purely
universal Horn frame conditions. Ostropolski-Nalewaja and
Lyon~\cite{OL2024} proved decidability for extensions of $\mathsf{K}$
by a finite set of quasi-density axioms
\(  \Diamond^k p\rightarrow\Diamond^n p\), for $0<k<n$.
Their frame correspondents require every pair of worlds connected by an
$R$-path of length $k$ to be connected by an $R$-path of length $n$.
These requirements have natural first-order descriptions with
existentially quantified intermediate worlds. The authors used
disjunctive existential rules and finite encodings of potentially
infinite models, called templates, to obtain an $\mathsf{EXPSPACE}$
decision procedure. This supplied a further decidable family of
elementary modal logics whose treatment required techniques beyond the
universal Horn analysis.

\textbf{Contribution.} Michaliszyn and Otop~\cite{MO12} and
Michaliszyn, Otop, and Kiero\'nski~\cite{MOK2015} asked whether the
decidable and undecidable cases among universal first-order frame
conditions could be distinguished algorithmically. We answer this
question for local satisfiability by proving that no such algorithm
exists. Theorem~\ref{thm:main} first establishes the result for universal
sentences over $\{R,=\}$. Section~\ref{sec:equality-elimination} strengthens
it to equality-free universal sentences.

The proof separates two cases of the domino problem. From a finite
system $\Dom$ we construct a universal frame condition $\Phi_{\Dom}$
encoding colors on ordered pairs of worlds. 
Compatibility along walks through proper edges between complete grid worlds turns every sufficiently long such walk into a tiling of a large finite square.
If $\Dom$
does not tile $\N^2$, the resulting uniform bound permits a finite-model
construction and yields decidability of $\Phi_{\Dom}$-SAT. If $\Dom$
does tile $\N^2$, its tiling supports a grid model with the required
color encoding. Modal formulas can then encode an arbitrary fresh
domino system $\Tiles$, making $\Phi_{\Dom}$-SAT undecidable while
$\Dom$ remains fixed. An algorithm separating these two cases would
therefore decide the domino problem.

\section{Preliminaries}
\subsection{Kripke structures and frames}
Kripke structures are tuples of the form $\naw{ W, R, \pi }$, where $\naw{ W, R }$ is a directed graph called a \emph{frame}, and $\pi$ is a labelling function that assigns to each element of $W$ a set of propositional variables which are true at this world. 
Members of $W$ are called worlds.

We say that a Kripke structure  $\naw{ W, R, \pi }$ is \emph{based} on the frame $\naw{ W, R }$. For a given class of frames $\klasa$, we say that a structure is $\klasa$-based if it is based on some frame from $\klasa$. 
We will use calligraphic letters ${\cal M}, {\cal N}$ to denote frames and Fraktur letters $\mathfrak{M}, \mathfrak{N}$ to denote structures. To keep the notation light, we use the notation $\naw{ {\cal M}, \pi}$, where ${\cal M} = \naw{ W, R }$ is a frame, to denote the structure $\naw{ W, R, \pi }$.

As we work with both first-order logic and modal logic, to distinguish them in our notation we denote first-order formulae with Greek capital letters, and modal formulae with Greek small letters. We assume that the reader is familiar with first-order logic and propositional logic.
For standard background on first-order structures and modal logic, see Hodges~\cite{hod93} and Blackburn, de Rijke, and Venema~\cite{modal}, respectively.

\subsection{Modal logic}
Modal logic extends propositional logic with the operator $\Diamond$ and its dual $\Box$.
The syntax of modal logic is given by the BNF \(
\varphi ::= p \mid \neg \varphi \mid \varphi \vee \varphi  \mid \Diamond \varphi \)
where $p$ is a propositional variable. 

The semantics of modal logic is defined recursively as follows.
For a model $\str M=\naw{W,R,\pi}$ and a world $w$, the relation
$\str M,w\models\varphi$ is specified by these clauses:
\begin{enumerate}[(i)]
\item $\str M,w\models p$ iff $p$ is a propositional variable and $p\in\pi(w)$.
\item $\str M,w\models\neg\psi$ iff $\str M,w\not\models\psi$.
\item $\str M,w\models\varphi_1\vee\varphi_2$ iff
      $\str M,w\models\varphi_1$ or $\str M,w\models\varphi_2$.
\item $\str M,w\models\Diamond\psi$ iff there is a world $v\in W$
      such that $(w,v)\in R$ and $\str M,v\models\psi$.
\end{enumerate}

We use the abbreviation $\Box \varphi$ to denote $\neg \Diamond \neg \varphi$ and $\Rightarrow$, $\land$, $\bot$, $\top$ are defined as usual.

By $|\varphi|$ we denote the length of $\varphi$. 
We say that a structure ${\str M}$ is a \emph{model} of $\varphi$ if there is a world $w$ such that ${\str  M}, w \models \varphi$.
For a given class of frames $\klasa$, we say that a formula $\varphi$ is ${\klasa}$-\emph{satisfiable} if there exists a $\klasa$-based model of $\varphi$.

\subsection{Decision problems}
For a first-order sentence $\Phi$ over the signature
$\{R,=\}$, let $\klasa_\Phi$ be the class of frames
satisfying $\Phi$.
Equality belongs to the first-order frame language; the modal
language remains plain unimodal propositional logic. 
For example, first-order functionality,
$\forall xyz\,(R(x,y)\land R(x,z)\Rightarrow y=z)$, corresponds to
frame validity of $\Diamond p\Rightarrow\Box p$. 

We define the \emph{satisfiability problem} \sat{\Phi} for modal logic \emph{over} $\kp$ as the question whether a given modal formula is $\kp$-satisfiable. 

We use the domino problem to separate the decidable and undecidable
cases of modal satisfiability. A \emph{domino system} is a tuple
${\cal D}=(D,D_H,D_V)$, where $D$ is a finite nonempty set of pieces
(colors), and $D_H,D_V\subseteq D\times D$ specify the permitted
horizontal and vertical adjacencies.
We say that $\cal D$ \emph{tiles} $\N \times \N$  if
there exists a function $t:\N \times \N \rightarrow { D}$ 
such that $\forall i,j \in \N$ we have $(t(i,j),t(i+ 1,j)) \in D_H$ and $(t(i,j),t(i,j+1)) \in D_V$.

The following lemma comes from \cite{Berger66,GK72}.

\begin{lemma} The following problem is undecidable: \label{doomino}
For a given domino system $\cal D$ determine if  $\cal D$ tiles $\N \times \N$.
\end{lemma}

\section{Undecidability of decidability}
In this section, we show that the following problem is undecidable.
We write $\SAT(\Phi)$ for the set of modal formulas satisfiable
over $\klasa_\Phi$; its membership problem is denoted by \sat{\Phi}.

\problem{Metaproblem}{Universal first-order sentence $\Phi$}{Is \sat{\Phi} decidable?}

\begin{theorem}\label{thm:main}
The metaproblem is undecidable.
\end{theorem}

We start with a proof with equality, then in section \ref{sec:equality-elimination} we show how to eliminate equality.

Fix a finite nonempty domino system $\Dom=(D,D_H,D_V)$, with
$D=\{c_1,\ldots,c_\ell\}$. The reduction constructs $\Phi_{\Dom}$ so
that $\Phi_{\Dom}$-SAT is decidable exactly when $\Dom$ does not tile
$\N^2$. We first explain the objects used to encode pair colors.
A world with a self-loop is called \emph{main}; a world without one
is \emph{auxiliary}. A main world is \emph{complete} when it points
to the start of an auxiliary chain with $\ell$ edges. The start is
denoted $z_0^x$, and the remaining positions are denoted
$c_1^x,\ldots,c_\ell^x$, as in the following drawing.

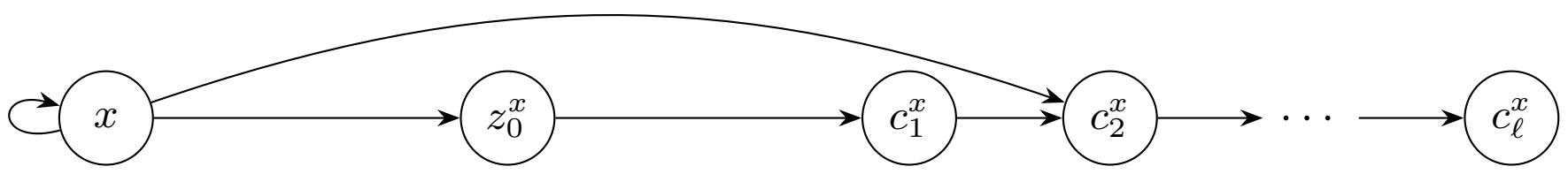
\begin{center}
\begin{tikzpicture}[
 vertex/.style={circle,draw,inner sep=1.2pt,minimum size=7mm,font=\small},
 >=Stealth,x=15mm,y=8mm]
 \node[vertex] (x) at (-2,0) {$x$};
 \node[vertex] (s) at (0,0) {$z_0^x$};
 \node[vertex] (c1) at (2,0) {$c_1^x$};
 \node[vertex] (c2) at (3,0) {$c_2^x$};
 \node (cd) at (4,0) {$\dots$};
 \node[vertex] (c4) at (5,0) {$c_\ell^x$};
 \draw[->] (x) edge[loop left,looseness=7] (x);
 \draw[->] (x) -- (s);
 \draw[->] (s) -- (c1);
 \draw[->] (c1) -- (c2);
 \draw[->] (c2) -- (cd);
 \draw[->] (cd) -- (c4);
 \draw[->] (x) to[out=19,in=161] (c2);
\end{tikzpicture}
\end{center}
{The chain is owned by $x$ through the edge to $z_0^x$.
A complete world with no edge to a color position of its own chain
has type $B$; one with exactly one such edge has type $G$.
The axioms allow no other case. Worlds of type $G$ will carry the
grid, and a world of type $B$ will make one local box apply to the
whole region used for the grid. Universal conditions constrain
these gadgets when they exist; they do not require every main
world to own a full chain.

The key encoding assigns a color to each ordered pair of $G$-worlds:
$C(x,y)=c_a$ means that $x$ points to position $c_a^y$ on the chain
owned by $y$. 
Whether an own-chain color is selected determines whether the complete world has type $B$ or $G$. The drawing displays the diagonal choice
$c_2$; it does not display every possible edge from $x$.
Let $S$ consist of edges between distinct $G$-worlds. The conditions
on pair colors enforce horizontal compatibility when the first
argument follows an $S$-edge and vertical compatibility when the
second argument follows an $S$-edge.

\emph{Non-tiling case.} An $S$-walk $a_0S\cdots Sa_{n-1}$ would
define a tiling of an $n\times n$ square by $t(i,j)=C(a_i,a_j)$,
even if worlds repeat. If $\Dom$ does not tile $\N^2$, some square
has no tiling at all. Consequently all such walks have a length
bound depending only on $\Dom$. Selecting finitely many modal
types and witnesses under this bound yields the finite-model
property. Searching for finite models and first-order refutations in parallel decides $\Phi_{\Dom}$-SAT.

\emph{Tiling case.} A tiling of $\Dom$ supplies the pair colors of
an actual grid frame with attached chains. For the fixed condition
$\Phi_{\Dom}$, a modal input encodes a fresh system $\Tiles$.
Its labels require two different successors at each grid world;
the frame condition makes those successors unique and forces
horizontal and vertical steps to commute. A $B$-world propagates
the local requirements throughout the chosen region. This yields
a map from $\N^2$ respecting the two steps and therefore a tiling
of $\Tiles$. Conversely, a tiling of $\Tiles$ supplies the modal
labels of the canonical grid model. Only $\Tiles$ varies in this
second reduction; $\Dom$ remains fixed.

}\subsection{Construction of first-order formula}

Fix $\Dom=(D,D_H,D_V)$ with $D=\{c_1,\ldots,c_\ell\}$ and
$\ell=|D|\geq1$. We now define $\Phi_{\Dom}$ formally.

Let \( \delta(x):=R(x,x)\) and \( \alpha(x):=\neg R(x,x)
\).

For readability, we omit universal quantifiers and take the universal
closure of each displayed condition. The sentence $\Phi_{\Dom}$ is
a conjunction of several such conditions. The first four are:
\begin{align}
 \alpha(x)\land R(x,y)&\ \Rightarrow\ \alpha(y),
 \label{eq:aux-start}\\
 \alpha(x)\land R(x,y)\land R(x,z)&\ \Rightarrow\ y=z,\\
 \alpha(x)\land\alpha(y)\land R(x,z)\land R(y,z)
 &\ \Rightarrow\ x=y,\\
 \neg\left(\bigwedge_{i=0}^{\ell+1}\alpha(z_i)
       \land\bigwedge_{i=0}^{\ell}R(z_i,z_{i+1})\right)&.
 \label{eq:aux-end}
\end{align}
Notice that formula \eqref{eq:aux-end} also excludes auxiliary cycles. 
The auxiliary part is a disjoint union of
finite directed chains with at most $\ell$ edges and has no edges to
main worlds.

For $\bar z=(z_0,\ldots,z_\ell)$ let
\(
 \Own(x,\bar z):=\delta(x)\land R(x,z_0)
 \land\bigwedge_{i=0}^{\ell}\alpha(z_i)
 \land\bigwedge_{i=0}^{\ell-1}R(z_i,z_{i+1}).
\)
Formula $\Phi_{\Dom}$ includes the following conjuncts (once again universal quantifiers are omitted):
\begin{align}
 \Own(x,\bar z)\land\Own(y,\bar z)&\ \Rightarrow\ x=y,
 \label{eq:owner}\\
 \Own(x,\bar z)\land\Own(x,\bar u)&\ \Rightarrow\ z_0=u_0.
 \label{eq:chainunique}
\end{align}
A main world $x$ is \emph{complete} precisely when
$\exists\bar z\,\Own(x,\bar z)$ holds: $x$ points to an auxiliary
chain with $\ell$ edges. This defines a class of worlds semantically
and adds no existential conjunct to $\Phi_{\Dom}$.
A complete world $x$ has a unique full chain, denoted
$\bar z^x=(z_0^x,\ldots,z_\ell^x)$. Full chains owned by different
worlds are disjoint: a full auxiliary chain cannot have an auxiliary
predecessor or a further successor, and the degree conditions prevent
chains from branching or merging. We will write $c_a^x:=z_a^x$ for
$1\leq a\leq\ell$. 

\begin{lemma}\label{lem:complete-test}
In every frame satisfying \eqref{eq:aux-start}--\eqref{eq:chainunique},
the formula
\(\beta=\Box^{\ell+1}\bot\) 
holds exactly at auxiliary worlds, and the formula
\( \gamma=\neg\beta\land\Diamond(\beta\land\Diamond^\ell\top)
\) holds exactly at complete worlds.
\end{lemma}
\begin{proof}
No auxiliary world starts a walk of length $\ell+1$. Main
worlds have self-loops. Thus $\beta$
recognizes auxiliary worlds. An auxiliary successor satisfying
$\Diamond^\ell\top$ starts a full auxiliary chain. For a main world,
having such a successor is exactly completeness.
\end{proof}

{
We introduce the abbreviations before specifying how they are
expanded into universal first-order sentences.
For complete worlds, put
\[
 B(x):=\bigwedge_{a=1}^{\ell}\neg R(x,c_a^x),
 \qquad G(x):=\bigvee_{a=1}^{\ell}R(x,c_a^x).
\]
For a finite list, $\One(\theta_1,\ldots,\theta_m)$ means that exactly
one entry is true, i.e., 
\(
 \left(\bigvee_{a=1}^{m}\theta_a\right)
 \land\bigwedge_{1\leq a<b\leq m}\neg(\theta_a\land\theta_b).
\)
We also use $\One$ for a finite set in any order. Define the edge
abbreviations
\[
 \begin{aligned}
 E(x,y)&:=R(x,y)\land x\ne y,\\
 U(x,y)&:=E(x,y)\land B(x)\land G(y),\\
 S(x,y)&:=E(x,y)\land G(x)\land G(y).
 \end{aligned}
\]
Thus $E$ is a proper edge, $U$ goes from a $B$-world to a $G$-world,
and $S$ is a proper edge between $G$-worlds. Finally,
$C(x,y)=c_a$ abbreviates $R(x,c_a^y)$; the pair-color axiom below
will ensure uniqueness when $x,y$ both have type $G$.

\subsection{Types from diagonal colors}

All conditions on complete worlds use the following guard convention.
For each displayed quantifier-free condition $\psi(x_1,\ldots,x_m)$
written with these abbreviations, include its guarded expansion
\begin{equation}\label{eq:guard}
 \forall x_1\cdots x_m\,\forall\bar z^1\cdots\forall\bar z^m
 \left(\bigwedge_{i=1}^m\Own(x_i,\bar z^i)
                 \ \Rightarrow\ \psi^*\right).
\end{equation}
Each $\bar z^i$ has $\ell+1$ variables. The formula $\psi^*$ is
obtained from that particular condition $\psi$ by expanding
$B,G,E,U,S,\One$, replacing $c_a^{x_i}$ by $z_a^i$, and replacing
$C(x_i,x_j)=c_a$ by $R(x_i,z_a^j)$. Membership in $D_H$ or $D_V$
is expanded into a finite disjunction. For instance,
\[
 (C(x_i,x_j),C(x_p,x_q))\in D_H
 \quad\text{expands to}\quad
 \bigvee_{(c_a,c_b)\in D_H}
 \bigl(R(x_i,z_a^j)\land R(x_p,z_b^q)\bigr).
\]
Different world variables may have the same value. Uniqueness of the
owned full chain makes the interpretation consistent in this case.

The first guarded condition permits at most one own-chain color:
\begin{equation}\label{eq:diagonal}
 \bigwedge_{1\leq a<b\leq\ell}
       \neg\bigl(R(x,c_a^x)\land R(x,c_b^x)\bigr).
\end{equation}
Thus each complete world has exactly one of the types $B,G$, with
type $G$ meaning exactly one own-chain color. For every pair of
complete worlds we also include
\begin{equation}\label{eq:paircolor}
 G(x)\land G(y)\ \Rightarrow\
        \One\bigl(R(x,c_1^y),\ldots,R(x,c_\ell^y)\bigr).
\end{equation}
This makes $C(x,y)$ a uniquely defined color for $G$-pairs, including
the case $x=y$.

}\subsection{The universal conditions on complete worlds}
In this subsection all variables range over complete worlds,
as formalised by \eqref{eq:guard}.

The next conditions to be added to $\Phi_{\Dom}$ have separate roles. Condition
\eqref{eq:allowed} permits proper complete edges only from $B$ to $G$
or from $G$ to $G$. Condition \eqref{eq:rootclosed} makes the region
seen from a $B$-world closed under grid steps, so a single box can
impose all local grid requirements. Conditions \eqref{eq:horizontal}
and \eqref{eq:vertical} propagate pair colors in their first and
second arguments, respectively, turning walks into finite tilings.
Condition \eqref{eq:degree} bounds the number of $S$-successors by
two; modal labels will require two different witnesses and hence
make them unique. Finally, \eqref{eq:intersection} will force the
successor pairs to meet, allowing coordinate labels to identify
the horizontal--vertical and vertical--horizontal endpoints.
All these conditions constrain existing edges and witnesses.

We add to $\Phi_{\Dom}$ the following conjuncts (adjusted by \eqref{eq:guard}):
\begin{align}
 E(x,y)&\ \Rightarrow\ U(x,y)\lor S(x,y),
 \label{eq:allowed}\\
 U(r,x)\land S(x,y)&\ \Rightarrow\ U(r,y),
 \label{eq:rootclosed}\\
 S(x,x')\land G(y)&\ \Rightarrow\
       (C(x,y),C(x',y))\in D_H,
 \label{eq:horizontal}\\
 S(y,y')\land G(x)&\ \Rightarrow\
       (C(x,y),C(x,y'))\in D_V.
 \label{eq:vertical}\\
\label{eq:degree}
 S(x,y_1)\land S(x,y_2)\land S(x,y_3)
&\ \Rightarrow\ y_1=y_2\lor y_1=y_3\lor y_2=y_3.
\end{align}

Finally, $\Phi_{\Dom}$ requires that if a world has two distinct $S$-successors $y,z$,
and each of them has two distinct $S$-successors,
then those two pairs must share a world.

\begin{equation}\label{eq:intersection}
 \begin{aligned}
 &S(x,y)\land S(x,z)\land y\ne z
 \land S(y,u_1)\land S(y,u_2)\land u_1\ne u_2
 \land S(z,v_1)\land S(z,v_2)\land v_1\ne v_2\\
 &\quad\Rightarrow
       u_1=v_1\lor u_1=v_2\lor u_2=v_1\lor u_2=v_2.
 \end{aligned}
\end{equation}
Together with \eqref{eq:degree}, this means that whenever two distinct \(S\)-successors of the same world each have two \(S\)-successors, their successor sets intersect.
The axiom is universal: it requires one of the displayed
equalities to hold, rather than asserting the existence
of an additional world.
Later, modal labels will identify this shared world
as the common endpoint of a horizontal step followed
by a vertical step and a vertical step followed
by a horizontal step.

Precisely, $\Phi_{\Dom}$ is the conjunction of
\eqref{eq:aux-start}--\eqref{eq:chainunique} and the guarded
expansions, according to \eqref{eq:guard}, of each individual
condition \eqref{eq:diagonal}--\eqref{eq:intersection}.
For each such condition, $\psi^*$ denotes its expanded version.
This finite conjunction is effectively constructible from $\Dom$,
uses only $R$ and equality, and can be written with a single
universal prefix after renaming bound variables.

\begin{figure}[tb]
\centering
\begin{tikzpicture}[
  >=Stealth,
  x=1cm,
  y=1cm,
  world/.style={
    circle,
    draw,
    minimum size=8mm,
    inner sep=1pt,
    font=\small
  },
  auxiliary/.style={
    circle,
    draw,
    minimum size=8mm,
    inner sep=1pt,
    font=\small
  },
  edge/.style={->,semithick},
  color edge/.style={
    ->,
    semithick,
    dashed
  },
  edge label/.style={
    font=\scriptsize,
    fill=white,
    inner sep=1.5pt
  }
]

% Main worlds.
\node[world] (b) at (0,0) {$b$};
\node[world] (x) at (4,0) {$x$};
\node[world] (y) at (8,0) {$y$};

% Owned auxiliary chains.
\node[auxiliary] (b0) at (0,-1.25) {$z_0^b$};
\node[auxiliary] (b1) at (0,-2.50) {$c_1^b$};
\node[auxiliary] (b2) at (0,-3.75) {$c_2^b$};

\node[auxiliary] (x0) at (4,-1.25) {$z_0^x$};
\node[auxiliary] (x1) at (4,-2.50) {$c_1^x$};
\node[auxiliary] (x2) at (4,-3.75) {$c_2^x$};

\node[auxiliary] (y0) at (8,-1.25) {$z_0^y$};
\node[auxiliary] (y1) at (8,-2.50) {$c_1^y$};
\node[auxiliary] (y2) at (8,-3.75) {$c_2^y$};

% Self-loops.
\draw[edge] (b) edge[loop above,looseness=7] (b);
\draw[edge] (x) edge[loop above,looseness=7] (x);
\draw[edge] (y) edge[loop above,looseness=7] (y);

% Edges between distinct complete worlds.
\draw[edge] (b) -- node[edge label,above] {$U$} (x);
\draw[edge] (x) -- node[edge label,above] {$S$} (y);
\draw[edge] (b) to[bend left=30]
  node[edge label,above,pos=.5] {$U$} (y);

% Ownership edges and auxiliary-chain edges.
\draw[edge] (b) -- (b0);
\draw[edge] (b0) -- (b1);
\draw[edge] (b1) -- (b2);

\draw[edge] (x) -- (x0);
\draw[edge] (x0) -- (x1);
\draw[edge] (x1) -- (x2);

\draw[edge] (y) -- (y0);
\draw[edge] (y0) -- (y1);
\draw[edge] (y1) -- (y2);

% Diagonal colors: C(x,x)=C(y,y)=c_1.
\draw[color edge]
  (x) to[out=215,in=180,looseness=1.15]
  node[edge label,left,pos=.55] {$C(x,x)$} (x1);

\draw[color edge]
  (y) to[out=325,in=0,looseness=1.15]
  node[edge label,right,pos=.55] {$C(y,y)$} (y1);

% Off-diagonal colors: C(x,y)=C(y,x)=c_2.
\draw[color edge]
  (x) to
  node[edge label,above right,pos=.65] {$C(x,y)$} (y2);

\draw[color edge]
  (y) to
  node[edge label,above left,pos=.65] {$C(y,x)$} (x2);

% Structural types.
\node[font=\small] at (0,-4.45) {$B(b)$};
\node[font=\small] at (4,-4.45) {$G(x)$};
\node[font=\small] at (8,-4.45) {$G(y)$};

\end{tikzpicture}

\caption{
A finite frame satisfying $\Phi_{\Dom}$ for
$D=\{c_1,c_2\}$ and
$D_H=D_V=\{(c_1,c_2),(c_2,c_1)\}$.
All arrows represent $R$-edges; dashed arrows select
pair colors, and their targets determine the selected colors.
Here $C(x,x)=C(y,y)=c_1$ and
$C(x,y)=C(y,x)=c_2$.
The world $b$ has no diagonal color and is therefore a $B$-world.
The edge $xSy$ satisfies both color-compatibility conditions.
}
\label{fig:complete-frame}
\end{figure}
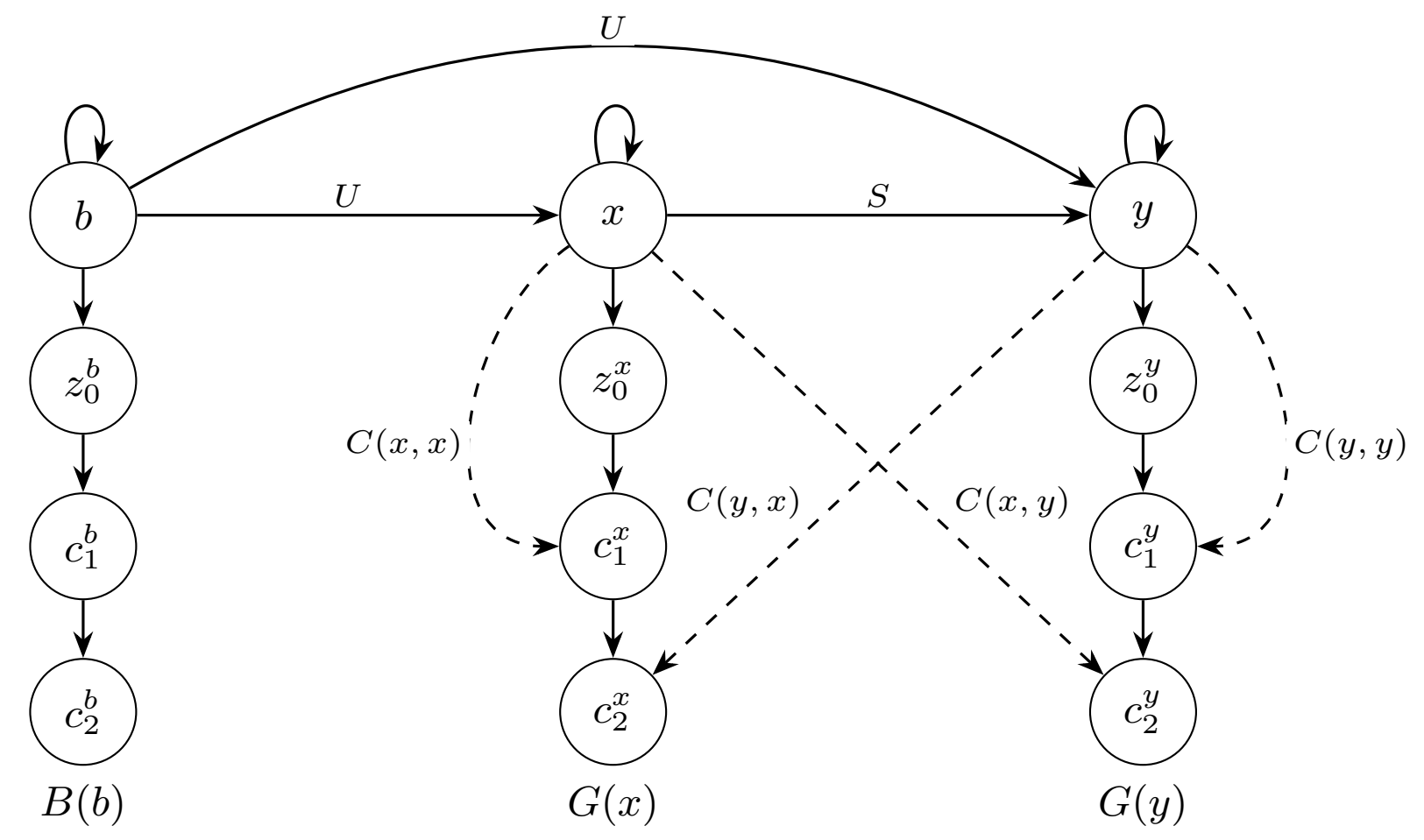

Figure~\ref{fig:complete-frame} illustrates the encoding
on a finite frame with three complete worlds.

\subsection{The non-tiling case}
The \emph{complete part} of a frame is the subframe induced by the complete worlds.

\begin{lemma}\label{lem:walkbound}
If $\Dom$ cannot tile an $n\times n$ square, no model of
$\Phi_{\Dom}$ has an $S$-walk $a_0Sa_1S\cdots Sa_{n-1}$ among
complete $G$-worlds, and every simple $R$-path in its
complete part has at most $n$ edges.
\end{lemma}
\begin{proof}
Such a walk would define $t(i,j)=C(a_i,a_j)$ for $0\leq i,j<n$.
Condition \eqref{eq:horizontal} gives horizontal compatibility and
\eqref{eq:vertical} gives vertical compatibility. This is a forbidden
square tiling. 
By \eqref{eq:allowed}, a simple path between complete worlds consists
of $S$-steps, possibly preceded by one $U$-step. No edge between
distinct complete worlds leads to a $B$-world. Self-loops do not
extend simple paths. The stated bound $n$ is therefore sufficient.
\end{proof}

\begin{lemma}
\label{lem:finite}
Let $\mathfrak M$ be a model over $\Phi_{\Dom}$. Suppose there is
$K\in\N$ such that every sequence $a_0Ra_1R\cdots Ra_m$ of pairwise
distinct complete worlds of $\mathfrak M$ satisfies $m\leq K$.
Then every modal formula satisfied in $\mathfrak M$ has a finite
model over $\Phi_{\Dom}$.
\end{lemma}

The bound in this lemma concerns one model. In the application,
there is a bound uniform over all models of $\Phi_{\Dom}$: if
$\Dom$ does not tile $\N^2$, the finitely branching tree of finite
square tilings has bounded height by K\"onig's lemma. Hence some
$n$, depending only on $\Dom$, admits no $n\times n$ tiling, and
Lemma~\ref{lem:walkbound} makes $K=n$ sufficient in every model.

\begin{proof}
Let $\mathfrak{M},w\models\varphi$. Eliminate boxes by
$\Box\psi\equiv\neg\Diamond\neg\psi$, and let $\Gamma$ be the set
of subformulas of the resulting formula. We define the \emph{modal type} of a world $a$ as
$
 \tp(a):=\{\psi\in\Gamma:\mathfrak{M},a\models\psi\}.
$
Let $m=|\Gamma|$. There are at most $2^m$ such types. 

If $w$ is auxiliary, its generated submodel is a finite chain tail by definition of $\Phi_{\Dom}$. Let us assume that $w$ is a main world.

We start by selecting complete worlds. There may be infinitely many such worlds, and a complete world may have infinitely many immediate successors. 
We address this by selecting one representative for every type and one witness for every needed subformula, as follows.

Let $A_0$ contain one representative of every modal type
realized by a complete world, choosing $w$ as the
representative of its type if $w$ is complete.
Thus $|A_0|\leq 2^m$.

For every complete world $a$ and every
$\Diamond\psi\in\Gamma$ true at $a$, fix an original
witness $b$ with $aRb$ and $\mathfrak M,b\models\psi$.
Let $A$ be the least set containing $A_0$ and closed
under these fixed witnesses whenever they are complete.

In the graph of selected witness edges between complete
worlds, every vertex has outdegree at most $m$.
Every world in $A$ is reachable from $A_0$ by a finite
path in this graph. Removing repetitions gives a simple
path, which has length at most $K$ because all its edges
are original $R$-edges between complete worlds. Hence
\(
 |A|\leq |A_0|\sum_{j=0}^{K}m^j
      \leq 2^m\sum_{j=0}^{K}m^j,
\)
so $A$ is finite.

{Next, choose one representative of every modal type realized
by an incomplete main world, and include $w$ if $w$ is an
incomplete main world. Call this finite set $F$.} 

We construct a finite model whose main worlds are copies of
$A\cup F$, together with finitely many auxiliary worlds.
We preserve the original propositional valuations and the
self-loop of every main world. For each world in $A$, we retain
its full owned chain, including its edges and valuations.
We retain all original edges between members of $A$ and all
original edges from members of $A$ to every position of every
retained owned chain.

For each $a\in A\cup F$ and each $\Diamond\psi\in\Gamma$ true
at $a$, choose an original witness $b$, using the previously
fixed choice when $a\in A$.
If $b$ is complete and $a\in A$, then $b\in A$, and the
required edge has already been retained.
If $b$ is complete and $a\in F$, add an edge to the initially
chosen complete representative of the modal type of $b$,
which belongs to $A$.
If $b$ is incomplete main, add an edge to its modal-type
representative in $F$.

If $b$ is auxiliary, consider its generated chain tail.
If this tail has length $\ell$, then $a$ is its complete
owner, so $a\in A$ and both the tail and the edge to its
beginning have already been retained.
Otherwise, attach a fresh disjoint copy of the entire tail
to $a$, preserving its edges and valuations.

{The complete worlds of the new model are exactly the copies
of $A$. A world of $F$ receives no edge to the start of a full
auxiliary chain, and every fresh tail has fewer than $\ell$ edges.
Thus no new complete world is created. Every world of $A$ retains
its unique full owned chain.

Auxiliary chains remain disjoint, have length at most $\ell$, and
have the prescribed unique ownership, so
\eqref{eq:aux-start}--\eqref{eq:chainunique} hold. Among worlds of
$A$, all original edges and all edges to positions of retained
owned chains are preserved exactly. Every active guarded instance
of \eqref{eq:diagonal}--\eqref{eq:intersection} therefore has the
same relational atoms and equality values as in the original
model, including its diagonal colors and the degree and
intersection disjunctions. Every other guarded instance has a
false ownership antecedent. Hence the frame satisfies $\Phi_{\Dom}$.}

Every retained or copied auxiliary tail is isomorphic, with its valuations, to its original generated submodel, so all modal formulas are preserved there. For each selected main world $a$, let $a'$ denote its copy. We prove by induction on $\psi\in\Gamma$ that $\mathfrak M',a'\models\psi$ iff $\mathfrak M,a\models\psi$. The atomic and Boolean cases are immediate.

For $\psi=\Diamond\eta$, every originally true diamond has a selected witness, which satisfies $\eta$ by induction or tail isomorphism. Conversely, every successor $u$ of $a'$ corresponds to an original successor $b$ of $a$: either $u$ copies a main world with the same original $\Gamma$-type as $b$, or its auxiliary tail is isomorphic to the original tail at $b$. Since $\eta\in\Gamma$, induction or tail isomorphism gives $\mathfrak M',u\models\eta$ iff $\mathfrak M,b\models\eta$. Thus no originally false diamond becomes true, completing the induction. In particular, $\mathfrak M',w'\models\varphi$.
\end{proof}

Finally, we show decidability.

\begin{proposition}\label{prop:decidable}
If $\Dom$ does not tile $\N^2$, then $\SAT(\Phi_{\Dom})$ is decidable.
\end{proposition}
\begin{proof}
By compactness there is an untileable $n\times n$ square, with
$n\geq2$. Lemmas \ref{lem:walkbound} and \ref{lem:finite} give the
finite model property over $\Phi_{\Dom}$.

Given $\varphi$, form the first-order sentence
$
 \Theta:=\Phi_{\Dom}\land\exists x\,\ST_x(\varphi).
$
The standard translation sends $p$ to a fresh unary predicate $Q_p(x)$,
commutes with Boolean connectives, and has the clause
$
 \ST_x(\Diamond\psi)=\exists y\,(R(x,y)\land\ST_y(\psi)),
$
using fresh bound variables. Interpretations of the unary predicates
are exactly propositional valuations. {Thus $\Theta$ is satisfiable if and only if
$\varphi\in\SAT(\Phi_{\Dom})$. More explicitly, $\Theta$ has a
finite first-order model if and only if $\varphi$ has a finite
Kripke model whose frame satisfies $\Phi_{\Dom}$: the standard
translation preserves the domain and accessibility relation,
and unary predicates encode precisely the propositional valuation.}

Interleave two effective searches. The first enumerates all structures
on $\{1,\ldots,m\}$ for $m=1,2,\ldots$ in the finite signature of
$\Theta$, checking which satisfy $\Theta$. The second enumerates
formal proofs of $\neg\Theta$ in a sound and complete effective
first-order calculus with equality. Proofs may use the full calculus,
not just the universal fragment. Soundness and completeness imply that 
 $\Theta\text{ has no model}$ if and only if $\vdash\neg\Theta$.
If $\Theta$ is satisfiable, the first search terminates by the finite
model property; otherwise the second search terminates. 
\end{proof}

\subsection{The tiling case}

The first step is to show that the pair-color encoding really works
on a two-dimensional grid, rather than only along a single walk.
The important point is that $\Dom$ does \emph{not} label individual
grid worlds: it colors \emph{ordered pairs} of grid worlds.  This
construction is independent of the fresh domino system $\Tiles$ used
later in the reduction.

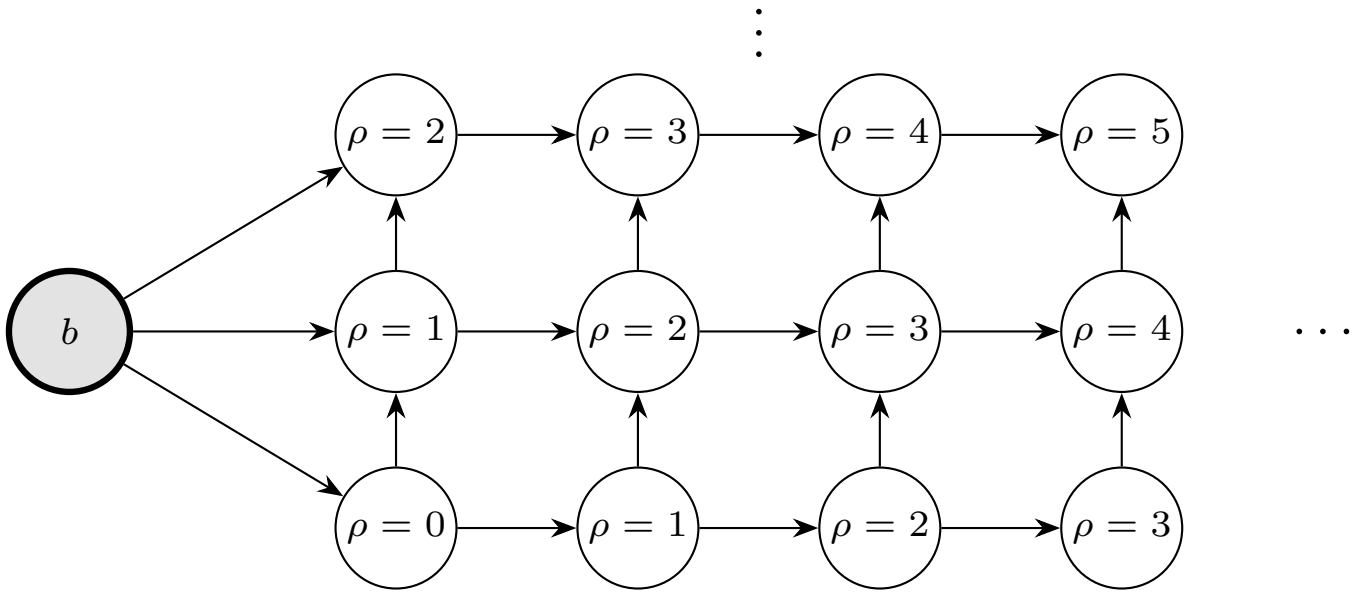
\begin{figure}[tb]
\centering
\begin{tikzpicture}[
 >=Stealth,
 x=16mm,y=13mm,
 gridworld/.style={circle,draw,thin,minimum size=8mm,inner sep=1.2pt,
                   fill=white,font=\scriptsize,align=center},
 marked/.style={gridworld,fill=black!11,very thick},
 note/.style={font=\scriptsize,align=center},
 arr/.style={->,thin},
 highlight/.style={->,very thick}
]
\foreach \i in {0,1,2,3}{
  \foreach \j in {0,1,2}{
    \pgfmathtruncatemacro{\rnk}{\i+\j}
    \node[gridworld] (g\i\j) at (\i,\j)
      {$\rho=\rnk$};
  }
}
\foreach \j in {0,1,2}{
  \foreach \i in {0,1,2}{
    \pgfmathtruncatemacro{\ip}{\i+1}
    \draw[arr] (g\i\j) -- (g\ip\j);
  }
}
\foreach \i in {0,1,2,3}{
  \foreach \j in {0,1}{
    \pgfmathtruncatemacro{\jp}{\j+1}
    \draw[arr] (g\i\j) -- (g\i\jp);
  }
}
\node[marked] (root) at (-1.35,1) {$b$};
\draw[arr] (root) -- (g00);
\draw[arr] (root) -- (g01);
\draw[arr] (root) -- (g02);

\node (dd) at (3.85,1) {\dots};
\node (dd) at (1.5,2.6) {\vdots};
\end{tikzpicture}
\caption{The grid frame used to realize the fixed system $\Dom$. A $B$-world $b$ sees the whole grid (only three of its edges are drawn), and each grid world $(i,j)$ carries the rank $\rho(i,j)=i+j$. The pair color assigned to any ordered pair of grid worlds is $C(u,v)=\tau(\rho(u),\rho(v))$. Since every step to the right or upward increases $\rho$ by $1$, the pair colors evolve according to the horizontal and vertical adjacency relations of the fixed tiling $\tau$. Self-loops, auxiliary chains, and color-encoding edges are omitted for readability.}
\label{fig:grid-realization}
\end{figure}

\begin{lemma}[Grid realization]\label{lem:grid-realization}
If $\Dom$ tiles $\N^2$, then there is a frame satisfying
$\Phi_{\Dom}$ with the following complete worlds: one $B$-world $b$
and a copy of $\N^2$ consisting of $G$-worlds.  The proper edges
between complete worlds are exactly the edges from $b$ to every
grid world and the horizontal and vertical successor edges:
\(
(i,j)\longrightarrow(i+1,j)\) and \(
(i,j)\longrightarrow(i,j+1).
\)
Every complete world has a disjoint full auxiliary chain, whose
start has no other predecessor.
\end{lemma}
\begin{proof}
Fix a tiling $\tau:\N^2\to D$ of $\Dom$.  Give $b$ and every
grid world a self-loop, connect $b$ to all grid worlds, and add
exactly the grid edges specified in the statement.  Attach to
each complete world $u$ its own auxiliary chain
$z_0^u R c_1^u R\cdots R c_\ell^u$, together with $uRz_0^u$.

It remains to choose the edges that encode colors.  Let 
$\rho(i,j):=i+j$ and $C(x,y):=\tau(\rho(x),\rho(y))
\text{ for all grid worlds }x,y.$

As usual, $C(x,y)=c_a$ is implemented by the edge $xRc_a^y$.
No other edges are added.  Thus the root has no diagonal color,
while every grid world has exactly one, and every ordered pair of
grid worlds receives exactly one color.

The reason this works on the entire grid is simple: \emph{both}
horizontal and vertical grid steps increase $\rho$ by one; see
Figure~\ref{fig:grid-realization}.  Hence, whenever $xSx'$, for
any grid world $y$ we have
\[
(C(x,y),C(x',y))
 =\bigl(\tau(\rho(x),\rho(y)),
        \tau(\rho(x)+1,\rho(y))\bigr)\in D_H.
\]
Likewise, whenever $ySy'$, for any grid world $x$ we have
\[
(C(x,y),C(x,y'))
 =\bigl(\tau(\rho(x),\rho(y)),
        \tau(\rho(x),\rho(y)+1)\bigr)\in D_V.
\]
These are precisely \eqref{eq:horizontal} and
\eqref{eq:vertical}.  The disjoint chains verify
\eqref{eq:aux-start}--\eqref{eq:chainunique}, and the uniqueness
of pair colors gives \eqref{eq:diagonal}--\eqref{eq:paircolor}.
Conditions \eqref{eq:allowed} and \eqref{eq:rootclosed} hold
because the only proper edges between complete worlds are the
specified grid edges and the edges from $b$ to the whole grid.
Every grid world has exactly two $S$-successors, giving
\eqref{eq:degree}; the successor sets of its right and upper
neighbors meet at the opposite corner of the grid square, giving
\eqref{eq:intersection}.  Therefore the frame satisfies
$\Phi_{\Dom}$.
\end{proof}

\begin{lemma}
\label{lem:tiling}
If $\Dom$ tiles $\N^2$, then
$\SAT(\Phi_{\Dom})$ is undecidable.
\end{lemma}

\begin{proof}[Proof idea]
Fix a tileable system $\Dom$. For each fresh domino system
$\Tiles=(T,T_H,T_V)$ we effectively construct a modal formula
$\varphi_{\Tiles}$ (see Appendix~\ref{sec:appendix-proofs}).
It requires a complete root with three pairwise distinct complete successors, all different from the root, so \eqref{eq:degree} forces the root to have type $B$.
By
\eqref{eq:rootclosed}, its complete proper successors form a nonempty
region of $G$-worlds closed under $S$-steps. A box at the root imposes
coordinate labels modulo three and $\Tiles$-tile labels throughout
this region. The coordinate labels require a horizontal and a vertical
successor at each world; \eqref{eq:degree} makes them unique, and
\eqref{eq:intersection} forces these two steps to commute. Hence the
region yields a (not necessarily injective) image of $\N^2$ carrying
a tiling of $\Tiles$. Conversely, any $\Tiles$-tiling can be placed
on the grid frame supplied by Lemma~\ref{lem:grid-realization},
producing a model of $\varphi_{\Tiles}$. The reduction is effective:
only $\Tiles$ varies, and the fixed $\Dom$-tiling is used to prove
existence of the grid frame, not to compute the modal formula.
Therefore the domino problem reduces to $\Phi_{\Dom}$-SAT.
The formal formula and the proof of both directions are given in
Appendix~\ref{sec:appendix-proofs}.
\end{proof}

Putting it together, we show:
\begin{proof}[Proof of Theorem~\ref{thm:main} with equality]
Given a domino system $\Dom$, construct $\Phi_{\Dom}$.
By Proposition~\ref{prop:decidable} and
Lemma~\ref{lem:tiling},
$\SAT(\Phi_{\Dom})\text{ is decidable}$ if and only if
$\Dom\text{ does not tile }\N^2$.
An algorithm for the metaproblem would therefore decide
the domino problem, contradicting Lemma~\ref{doomino}.
\end{proof}

\subsection{Eliminating equality}
\label{sec:equality-elimination}

We strengthen Theorem~\ref{thm:main} to equality-free universal frame
conditions over $\{R\}$. The following lemma transfers decidability by an effective
translation of modal formulas. Undecidability will be transferred separately.

A universal sentence is \emph{equality-positive} if equality occurs only
positively in negation normal form; occurrences of $R$ may have either
polarity. Define
\[
 \operatorname{Tw}(x,y):=\forall z\,
 \bigl((R(x,z)\leftrightarrow R(y,z))\land
       (R(z,x)\leftrightarrow R(z,y))\bigr).
\]
Worlds satisfying this formula are called \emph{twins}. In a frame
$\mathcal M$, write $u\sim v$ for $\mathcal M\models\operatorname{Tw}(u,v)$.
Let $\Phi^{\operatorname{Tw}}$ replace every equality atom in $\Phi$
by a copy of $\operatorname{Tw}$ with fresh bound variables.

\begin{lemma}[Equality elimination]\label{lem:equality-elimination-quotient}
For every universal equality-positive sentence $\Phi$ over $\{R,=\}$,
an equality-free universal sentence $\Psi\equiv\Phi^{\operatorname{Tw}}$
is effectively computable and satisfies:
\begin{enumerate}
\item For every frame $\mathcal M$,
$\mathcal M\models\Psi$ iff $\mathcal M/{\sim}\models\Phi$.
\item If $\mathcal F\models\Phi$ and a surjection
$f:\mathcal N\twoheadrightarrow\mathcal F$ satisfies
$R_{\mathcal N}(u,v)\iff R_{\mathcal F}(f(u),f(v))$,
then $\mathcal N\models\Psi$. In particular,
$\klasa_\Phi\subseteq\klasa_\Psi$.
\item There is an effective translation $\varphi\mapsto\varphi^\sharp$ such that
\(
 \varphi\in\operatorname{SAT}(\Psi)
\) iff \(
 \varphi^\sharp\in\operatorname{SAT}(\Phi).
\)
Consequently, decidability of $\Phi$-SAT implies decidability of $\Psi$-SAT.
\end{enumerate}
\end{lemma}

The translation in 3 is needed because twins may carry different
propositional valuations; the same modal formula need not be satisfiable
over both classes.

\begin{proof}
Equality positivity ensures that replacing equality by
$\operatorname{Tw}$ introduces universal quantifiers only in positive
positions. Renaming bound variables and moving these quantifiers to
the front gives $\Psi$.

For 1, twins form an equivalence relation and a congruence for $R$.
The quotient map $q$ therefore satisfies
\(
 R_{\mathcal M}(u,v)\iff R_{\mathcal M/{\sim}}(q(u),q(v))\) and
 \(\operatorname{Tw}(u,v)\iff q(u)=q(v).\)
Induction on first-order formulas, using surjectivity at quantifiers,
gives the claim. 

For 2, interpret equality in $\mathcal N$ as
$E_f(u,v)\iff f(u)=f(v)$. Pulling back along $f$ preserves the truth
of $\Phi$, and $E_f\subseteq\sim$. Enlarging this interpretation to
$\sim$ preserves truth by equality positivity. The identity map gives
the stated inclusion of frame classes.

For 3, fix a modal formula $\varphi$ and let $P$ be its set of
propositional letters.
The difficulty is that several twin worlds may be merged into
one world in the quotient, even though they have different
propositional valuations. We simulate these different twins
using fresh propositional letters $A_s$, one for each
$s\subseteq P$. The meaning of $A_s$ is that a twin with
valuation $s$ is available at the current world.
Several such markers may hold at the same world.

For each $s\subseteq P$, define a translation
$\eta\mapsto\eta^s$. It evaluates propositional letters
according to $s$ ($p^s$ holds iff $p \in s$) and commutes with Boolean connectives $(\psi_1 \land \psi_2)^s = \psi_1^s \land \psi_2^s$.

The key rule is
\[
(\Diamond\eta)^s :=
\Diamond\bigvee_{t\subseteq P}(A_t\land\eta^t).
\]
Here the choice of $s$ is forgotten when we pass through a
diamond. At the successor world we may choose any available
valuation $t$, independently of the valuation at the current
world. Finally, we allow any available valuation at the
initial world: 
\(
\varphi^\sharp :=
\bigvee_{s\subseteq P}(A_s\land\varphi^s).
\)

We show that this translation works in both directions.

First, suppose that $\varphi$ holds in a model over $\Psi$.
Take its quotient by the twin relation, which satisfies
$\Phi$ by part~1. At each quotient world, let $A_s$ hold
if some world in the corresponding twin class has valuation
$s$. Since twins have exactly the same successors, moving
from one twin to a successor does not depend on which twin
we started from. Thus the quotient, equipped with the
markers $A_s$, simulates all the original twins.
By induction on the formula, it satisfies
$\varphi^\sharp$.

Conversely, suppose that $\varphi^\sharp$ holds in a model
over $\Phi$. We reverse the simulation: replace each world
$v$ by one copy $(v,s)$ for every marker $A_s$ true at $v$.
The copy $(v,s)$ has precisely the propositional valuation
$s$. Every copy of $v$ is connected to every copy of each
successor of $v$. Thus a diamond can move to any available
copy of a successor, exactly as prescribed by the translation.

We discard worlds carrying no marker. 
Removing these worlds does not change the truth of any translated formula \(\eta^s\) at the remaining worlds, since every translated diamond explicitly requires its successor to carry some marker \(A_t\).
Since $\Phi$ is
universal, the remaining subframe still satisfies $\Phi$.
Part~2 then ensures that the frame consisting of all the
copies satisfies $\Psi$. A straightforward induction shows
that a copy $(v,s)$ satisfies $\eta$ exactly when $v$
satisfies $\eta^s$. Hence some copy of the initial world
satisfies $\varphi$.
The translation is effective, proving~3.
\end{proof}

\begin{proof}[Proof of Theorem~\ref{thm:main} without equality]
Fix a domino system $\Dom$. We first rewrite $\Phi_{\Dom}$ as an equivalent
equality-positive universal sentence. Under the guard convention~\eqref{eq:guard},
replace \eqref{eq:allowed} by \( R(x,y)\Longrightarrow\ x=y\lor G(y)\) and \eqref{eq:rootclosed}, by
\(
 B(r)\land R(r,x)\land R(x,y)
 \Longrightarrow\ R(r,y)\).
The first is equivalent to \eqref{eq:allowed}, since $B,G$ partition
complete worlds. Given this condition, the second is equivalent to
\eqref{eq:rootclosed}: if $x=r$ or $y=x$, its conclusion is already
in the premise; otherwise the first condition gives $G(x)$ and $G(y)$,
so \eqref{eq:rootclosed} applies. Conversely, the second condition
implies \eqref{eq:rootclosed}, whose premise already supplies
$B(r),G(y)$ and hence $r\ne y$.
In the remaining conditions, inequalities occur only in antecedents
and equalities only in consequents, so all equality occurrences are
positive in negation normal form. The guards are equality-free, and
$C(x,y)=c_a$ abbreviates an $R$-atom. Expanding the abbreviations
therefore yields an equality-positive universal sentence $\Phi_{\Dom}^+$
with $\klasa_{\Phi_{\Dom}^+}=\klasa_{\Phi_{\Dom}}$.

Let $\Psi_{\Dom}$ be the universal prenex form of
$(\Phi_{\Dom}^+)^{\operatorname{Tw}}$. This is equality-free, and the map
$\Dom\mapsto\Psi_{\Dom}$ is effective. If $\Dom$ does not tile $\N^2$,
Proposition~\ref{prop:decidable} gives decidability of $\Phi_{\Dom}$-SAT,
and part~(3) of Lemma~\ref{lem:equality-elimination-quotient} gives
decidability of $\Psi_{\Dom}$-SAT.

Suppose that $\Dom$ tiles $\N^2$. For a fresh domino system $\Tiles$,
take the modal formula $\varphi_{\Tiles}$ from Lemma~\ref{lem:tiling}, and let
$\widehat\varphi_{\Tiles}$ be obtained by simultaneously replacing every
propositional letter $p$ in $\varphi_{\Tiles}$ by $\Diamond p$. We claim
\(
 \widehat\varphi_{\Tiles}\in\operatorname{SAT}(\Psi_{\Dom})\)
iff \( \Tiles\text{ tiles }\N^2.
\)
For the forward implication, let
$\mathfrak M,x\models\widehat\varphi_{\Tiles}$ on a $\Psi_{\Dom}$-frame.
Its twin quotient $\mathcal Q$ satisfies $\Phi_{\Dom}$. Since $\Diamond p$
is constant on twin classes, the valuation
\(
 \mathfrak Q,[u]\models p
 \) iff \(
 \mathfrak M,u\models\Diamond p
\)
is well-defined. Since the quotient map is onto and preserves and
reflects edges, induction on the modal formula gives
$\mathfrak Q,[x]\models\varphi_{\Tiles}$. Lemma~\ref{lem:tiling}
then yields a tiling of $\Tiles$.

Conversely, if $\Tiles$ tiles $\N^2$, use the canonical model from
the converse part of Lemma~\ref{lem:tiling}. All propositional letters of
$\varphi_{\Tiles}$ may be taken to be false on auxiliary worlds.
Every main world $u$ has a private chain start $z_0^u$, whose only
predecessor is $u$: color edges target positions $1,\ldots,\ell$,
never position~$0$. Transfer each letter $p$ from $u$ to $z_0^u$,
setting it false everywhere else. For every world $v$, the new
truth value of $\Diamond p$ is exactly the old truth value of $p$.
Induction therefore gives $\widehat\varphi_{\Tiles}$ at the root under the
new valuation. The frame satisfies $\Phi_{\Dom}^+$ and hence $\Psi_{\Dom}$
by part~(2) of Lemma~\ref{lem:equality-elimination-quotient}.

The map $\Tiles\mapsto\widehat\varphi_{\Tiles}$ is effective, so $\Psi_{\Dom}$-SAT is
undecidable when $\Dom$ tiles $\N^2$. Consequently,
\(
 \Psi_{\Dom}\text{-SAT is decidable}
 \) iff \(\Dom\text{ does not tile }\N^2.
\)
Since $\Dom\mapsto\Psi_{\Dom}$ is effective, an algorithm for the stated
metaproblem would decide the domino problem, a contradiction.
\end{proof}

\section{Conclusion and future work}

We have shown that it is undecidable whether an equality-free
universal first-order sentence over a single binary relation
defines a class of frames with a decidable local modal
satisfiability problem. Thus no algorithm can distinguish
all decidable cases from undecidable ones, even under
these syntactic restrictions.

The reduction relies on two complementary properties
of the constructed frame conditions. Pair colors turn
$S$-walks among complete $G$-worlds into finite square
tilings. When the fixed domino system does not tile
the plane, this gives a uniform bound on the length
of such walks, leading to the finite model property
and a decision procedure. When the system does tile
the plane, the same frame condition admits a grid
model on which modal formulas can encode arbitrary
domino systems. These two cases establish the
undecidability of the metaproblem.

A natural question is whether the result extends to
global satisfiability. In that setting, grid conditions
could be imposed throughout the model, potentially
eliminating the need for the $B$-world and its
propagation conditions. However, the auxiliary-chain
encoding would need to be modified. The terminal
worlds of these chains generate singleton submodels,
which inherit both the universal frame condition
and any globally satisfied modal formula. This
prevents the present construction from enforcing
the intended grid behaviour through global
satisfiability.

Another open question concerns the metaproblem
for finite satisfiability. The finite model property
already gives decidability in the non-tiling case
of our reduction. However, undecidability in the
tiling case is established using an infinite grid
model, so the argument does not directly extend
to finite satisfiability. A different construction
would be needed to obtain such a result. The
distinction is particularly relevant because
ordinary and finite satisfiability can have
different decidability statuses even for a fixed
universal frame condition~\cite{MOW2018}.

\bibliography{all}{}

\appendix
\section{Proof of Lemma~\ref{lem:tiling}}\label{sec:appendix-proofs}
Fix a tileable system $\Dom$. For a fresh domino system
$\Tiles=(T,T_H,T_V)$, we construct a modal formula $\varphi_{\Tiles}$
that is satisfiable over $\Phi_{\Dom}$ exactly when $\Tiles$ tiles
$\N^2$. The system $\Dom$ remains fixed; the grid frame needed for
the converse direction is provided by Lemma~\ref{lem:grid-realization}.

We use fresh propositional variables $r,e_1,e_2,e_3$,
variables $q_{ij}$ for $i,j\in\{0,1,2\}$,
and variables $p_t$ for $t\in T$.
All indices of coordinate labels are calculated modulo $3$.
Recall that $\gamma$ recognizes complete worlds.
Define
\[
 B_3:=
 \bigwedge_{a=1}^3
 \Diamond\left(
   \gamma\land\neg r\land e_a
   \land
   \bigwedge_{\substack{1\le b\le3\\b\ne a}}\neg e_b
 \right).
\]
The witnesses of these three diamonds are pairwise distinct.
At a world satisfying $r$, they are also different from
that world.

Let $\chi_{\Tiles}$ be the conjunction of the following
formulas, where $i,j$ range over $\{0,1,2\}$ and $t$ ranges
over $T$:
\begin{gather}
 \One(\{q_{ij}:i,j\in\{0,1,2\}\})
 \land
 \One(\{p_t:t\in T\}),
 \label{eq:unique-modal}\\
 \bigwedge_{i,j}
 \left[
   q_{ij}\Rightarrow
   \left(
     \Diamond(\gamma\land\neg r\land q_{i+1,j})
     \land
     \Diamond(\gamma\land\neg r\land q_{i,j+1})
   \right)
 \right],
 \label{eq:modal-successors}\\
 \bigwedge_{i,j,t}
 \left[
   (q_{ij}\land p_t)\Rightarrow
   \Box\left(
     (\gamma\land\neg r\land q_{i+1,j})
     \Rightarrow
     \bigvee_{(t,u)\in T_H}p_u
   \right)
 \right],
 \label{eq:modal-horizontal}\\
 \bigwedge_{i,j,t}
 \left[
   (q_{ij}\land p_t)\Rightarrow
   \Box\left(
     (\gamma\land\neg r\land q_{i,j+1})
     \Rightarrow
     \bigvee_{(t,u)\in T_V}p_u
   \right)
 \right].
 \label{eq:modal-vertical}
\end{gather}
As usual, an empty disjunction denotes $\bot$.
Finally, let
\begin{equation}\label{eq:modal-input}
 \varphi_{\Tiles}:=
 r\land\gamma\land B_3
 \land
 \Box\left(
   \gamma\Rightarrow
   \bigl((r\land B_3)\lor(\neg r\land\chi_{\Tiles})\bigr)
 \right).
\end{equation}

Assume first that ${\str M},w\models\varphi_{\Tiles}$ and that
the frame of $\str M$ satisfies $\Phi_{\Dom}$.
By Lemma~\ref{lem:complete-test}, $w$ is complete.
If $w$ had type $G$, the formula $r\land B_3$ would give
three distinct complete successors other than $w$,
contrary to \eqref{eq:degree}.
Thus $w$ has type $B$.

Let $Z$ be the set of complete successors of $w$ other
than $w$. By $B_3$, this set is nonempty.
By \eqref{eq:allowed}, its elements have type $G$
and are exactly the $U$-successors of $w$.
The box in \eqref{eq:modal-input} implies that every
$z\in Z$ satisfies
$ (r\land B_3)\lor(\neg r\land\chi_{\Tiles}).
$
The first alternative is impossible at a $G$-world
by the same degree argument. Hence every $z\in Z$
satisfies $\neg r\land\chi_{\Tiles}$.

Condition \eqref{eq:rootclosed} makes $Z$ closed under
$S$-successors. Moreover, every complete successor of
a world $z\in Z$ belongs to $Z$: it is either $z$ itself
or, by \eqref{eq:allowed}, an $S$-successor of $z$.

Suppose that $z\in Z$ has label $q_{ij}$.
By \eqref{eq:modal-successors}, it has complete successors
with labels $q_{i+1,j}$ and $q_{i,j+1}$.
These successors belong to $Z$, so each has exactly
one coordinate label by \eqref{eq:unique-modal}.
The two required labels are distinct from one another
and from $q_{ij}$. Thus the witnesses are distinct
from one another and from $z$.
By \eqref{eq:degree}, they are exactly the two
$S$-successors of $z$.
Denote them by $h(z)$ and $v(z)$, respectively.
This defines total functions $h,v:Z\to Z$.

We show that $v(h(z))=h(v(z))$ for every $z\in Z$.
Let $z$ have label $q_{ij}$, and let $b=h(z)$
and $c=v(z)$.
The worlds $b,c$ are distinct $S$-successors of $z$,
and their respective pairs of distinct $S$-successors
are $h(b),v(b)$ and $h(c),v(c)$.
Applying \eqref{eq:intersection} gives
$
 \{h(b),v(b)\}\cap\{h(c),v(c)\}\ne\varnothing.
$
The labels of the first pair are
$q_{i+2,j}$ and $q_{i+1,j+1}$, while those of the
second pair are $q_{i+1,j+1}$ and $q_{i,j+2}$.
Modulo $3$, their only common label is $q_{i+1,j+1}$.
Since every world in $Z$ has exactly one coordinate
label, the only possible equality between these
pairs is $v(b)=h(c)$, as required.

Choose $z_0\in Z$ and define
$
 f(i,j):=h^i(v^j(z_0))
$ for $i, j \in \N$.
Commutativity yields
\[
 f(i+1,j)=h(f(i,j)),
 \qquad
 f(i,j+1)=v(f(i,j)).
\]
Define $\eta(i,j)$ to be the unique tile $t\in T$
such that ${\str M},f(i,j)\models p_t$.
By \eqref{eq:modal-horizontal} and
\eqref{eq:modal-vertical}, $\eta$ is a tiling of
$\Tiles$.
The map $f$ need not be injective.

Conversely, suppose that $\eta:\N^2\to T$ is a tiling
of $\Tiles$. Take the frame from Lemma~\ref{lem:grid-realization},
with root $b$. That lemma already verifies all conditions of
$\Phi_{\Dom}$, including the pair-color encoding of $\Dom$;
we only need to assign propositional labels.

Let $r$ hold exactly at $b$. At each grid world $(i,j)$, let
$q_{kl}$ hold exactly when $i\equiv k\pmod 3$ and
$j\equiv l\pmod 3$, and let $p_t$ hold exactly when
$\eta(i,j)=t$. Give three distinct grid worlds the three mutually
exclusive $e$-patterns required by $B_3$. All remaining
propositional letters are false, in particular at every auxiliary
world.

Every grid world satisfies $\neg r\land\chi_{\Tiles}$:
its unique coordinate and tile labels verify
\eqref{eq:unique-modal}; its two grid successors verify
\eqref{eq:modal-successors}; and the horizontal and vertical
adjacencies of $\eta$ verify \eqref{eq:modal-horizontal} and
\eqref{eq:modal-vertical}. The root satisfies
$r\land\gamma\land B_3$. Its complete successors are the root
itself, satisfying $r\land B_3$, and all grid worlds, satisfying
$\neg r\land\chi_{\Tiles}$. Its auxiliary successors do not
satisfy $\gamma$. Therefore the box in \eqref{eq:modal-input}
holds at $b$, and $\mathfrak M,b\models\varphi_{\Tiles}$.

For the fixed $\Dom$, the map $\Tiles\mapsto\varphi_{\Tiles}$ is effective.
Although the grid frame exists because $\Dom$ tiles $\N^2$,
its construction need not be effective and no tiling of $\Dom$
needs to be computed in the reduction. As the domino problem is
undecidable, $\SAT(\Phi_{\Dom})$ is undecidable.

\section{Use of AI tools}
An AI tool (GPT-6 Astra) was used during the design of this research project. In the course of several interactions, the tool proposed multiple lengthy proof candidates that are not included in this paper. Drawing on ideas that emerged during these interactions, we subsequently developed a substantially simpler proof, which is the one presented here.

The same AI tool was also used to assist with revisions  and language simplification, including improvements to clarity, concision, and style, and to generate Figure~\ref{fig:complete-frame}. All mathematical arguments and results presented in the paper were independently reviewed and verified by the author.

\end{document}